\relax
\documentclass[a4paper,12pt]{article}
\usepackage[italian]{babel}
\usepackage{amssymb,amsmath,amsfonts}
\usepackage{graphicx}
\usepackage{subfigure}
\renewcommand{\epsilon}{\varepsilon}
\renewcommand{\phi}{\varphi}
\newcommand{\reali}{{\mathbb{R}}}

\newcommand{\interi}{{\mathbb{Z}}}

\newcommand{\Rscr}{{\cal R}}
\def\corsivo{\sl}
\def\gt{>}
\def\lt{<}
\def\ge{\geq}
\def\:{'{\kern .03em}}

\def\dbiref#1{\cite{#1}}

\title{Su un\:estensione della teoria di Lagrange \\
per i moti secolari.}

\author{Antonio Giorgilli\footnote{Universit\`a degli Studi di Milano,
Dipartimento di Matematica,
Via Saldini 50, 20133\ ---\ Milano.\hfill\break
email: antonio.giorgilli@unimi.it.\hfill\break
email: marco.sansottera@unimi.it}{\ },
Ugo Locatelli\footnote{Universit\`a  degli Studi di Roma ``Tor
Vergata'', Dipartimento di Matematica,
Via della Ricerca Scientifica 1, 00133 Roma.\hfill\break
email: locatell@mat.uniroma2.it}{\ },
Marco Sansottera${}^*$}

\date{}

\begin{document}
\maketitle

\begin{abstract}
La teoria di Lagrange per i moti secolari delle eccentricit\`a ed
inclinazioni delle orbite planetarie si fondava su
un\:approssimazione, dettata in larga misura dalla complessit\`a dei
calcoli necessari, che consisteva nel considerare solo equazioni
lineari.  In questa memoria riprendiamo in considerazione i metodi di
Lagrange alla luce della teoria della stabilit\`a esponenziale di
Nekhoroshev.  Grazie agli algoritmi sviluppati negli ultimi anni e
alle tecniche di manipolazione algebrica possiamo tener conto anche
dei contributi non lineari alle equazioni.  Come applicazione
cerchiamo di determinare i tempi di stabilit\`a per il problema dei
tre corpi nel caso del Sole e dei due pianeti maggiori, Giove e
Saturno, mostrando che si possono ottenere risultati realistici,
ancorch\'e non ottimali.

\vskip 18pt 
Lagrange's theory for the secular motion of perihelia and nodes of the
planetary orbits was based on consideration of a linear approssimation
of the dynamical equations, compatible with the complexity of the
calculations.  We extend Lagrange's investigations in the light of
Nekhoroshev's theory of exponential stability.  Using effective
algorithms recently developed and computer algebra we investigate the
non linear problem.  We apply our methods to the problem of three
bodies in the Sun--Jupiter--Saturn case, thus showing that realistic
results, although not optimal, can be obtained.
\end{abstract}

\section{Introduzione}\label{sec:1}
Nel 1782, in due corpose memorie presentate all\:Accademia di Berlino,
Giuseppe Luigi Lagrange pubblica la versione estesa della sua teoria
sui moti secolari dei nodi e dei perieli dei pianeti, e conclude il
suo studio con l\:affermazione della stabilit\`a del Sistema Solare.
Si tratta, per quel tempo, di un risultato di notevole rilevanza,
soprattutto se si tiene conto che viene pubblicato in un periodo in
cui sono molto vive le discussioni sulle ``ineguaglianze secolari'',
in primis quella di Giove e Saturno, e sull\:effettiva possibilit\`a
di spiegare tutte le apparenti irregolarit\`a dei moti planetari sulla
base della gravitazione newtoniana.

Alla luce delle nostre conoscenze attuali, e con la sensibilit\`a
matematica del nostro tempo, quel risultato non appare completamente
rigoroso, in quanto frutto di approssimazioni la cui completa
validit\`a non \`e assicurata.  In questa memoria vogliamo illustrare
come ed in che senso si possa estendere la teoria di Lagrange tenendo
conto degli sviluppi recenti delle nostre conoscenze.

Il nostro metodo si ricollega in modo diretto a quello di Lagrange in
quanto prendiamo come riferimento le orbite circolari, studiando poi
l\:evoluzione delle eccentricit\`a e delle inclinazioni come
oscillazioni intorno ad un equilibrio.  L\:estensione rispetto al
metodo di Lagrange prende avvio dalla teoria della stabilit\`a
esponenziale, sviluppata in forma teorica da
Moser~\dbiref{Moser-55.1},
Littlewood~\dbiref{Littlewood-59}$\,$\dbiref{Littlewood-59.1} e
Nekhoroshev~\dbiref{Nekhoroshev-77}$\,$\dbiref{Nekhoroshev-79}.  Al
fine di applicare tale teoria, dobbiamo procedere allo sviluppo in
serie della perturbazione mutua di Giove e Saturno, tenendo conto anche
dei termini di secondo ordine nelle masse e dei termini non lineari
nelle eccentricit\`a.  Ci\`o risulta fattibile grazie alla
disponibilit\`a dei metodi di manipolazione algebrica al calcolatore,
che rendono calcolabili sviluppi in serie praticamente impossibili in
passato.  Questo aspetto viene discusso brevemente nel
paragrafo~\ref{sec:2}, in cui richiamiamo anche i punti essenziali del
metodo di Lagrange.

In un secondo tempo facciamo ricorso al metodo della forma normale di
Birkhoff rienunciato mediante algoritmi di calcolo esplicitamente
applicabili grazie alla manipolazione algebrica. Questi metodi vengono
richiamati nel paragrafo~\ref{sec:3}.  Infine mostriamo come si
possano ottenere stime di stabilit\`a su tempi lunghi sfruttando la
forma normale ad ordine finito; questo viene discusso nel
paragrafo~\ref{sec:4}.

Il paragrafo~\ref{sec:5} riporta i risultati dell\:applicazione del
nostro metodo al sistema Sole--Giove--Saturno.  Le conclusioni, che
discutiamo brevemente nel paragrafo~\ref{sec:6}, non
saranno completamente soddisfacenti: si arriva a garantire la
stabilit\`a solo sull\:arco di $10^7$ anni, nettamente inferiore
rispetto alla durata del Sistema Solare ed anche rispetto ai tempi
calcolati mediante integrazione diretta delle equazioni.  Si tratta
comunque di uno tra i migliori risultati oggi disponibili sulla base
dell\:applicazione dei metodi perturbativi ed in particolare mostra
come il meccanismo della stabilit\`a su tempi esponenziali possa ben
essere significativo almeno per i pianeti maggiori del nostro sistema.

\section{Il problema secolare}\label{sec:2}
Lo schema di sviluppo perturbativo che utilizziamo pu\`o considerarsi
come una riformulazione in ambito hamiltoniano dello schema di
Lagrange.  Questa parte del calcolo risulta alquanto laboriosa, ma
omettiamo i dettagli in quanto si tratta di un argomento classico che
si pu\`o trovare ad esempio nelle {\corsivo Le\c cons de M\'ecanique
C\'eleste} di Poincar\'e.  Partendo dall\:Hamiltoniana del problema
dei tre corpi si procede in modo classico introducendo le coordinate
eliocentriche, che consentono di eliminare il moto del baricentro, ed
effettuando la riduzione del momento angolare.  In tal modo si
eliminano 5 gradi di libert\`a, sicch\'e ci si riduce a considerare 4
sole coppie di coordinate canoniche.

Si considerano poi gli {\corsivo elementi orbitali}, ossia il semiasse
maggiore $a$, l\:eccentricit\`a $e$, l\:inclinazione $\iota$,
l\:anomalia media $\ell$, l\:argomento del perielio $\omega$ e la
longitudine del nodo $\Omega$, ed in termini di questi si introducono
le variabili di Poincar\'e definite come
$$
\vcenter{\openup1\jot\halign{
\hfil${#}$
&${#}\>,\quad\ $\hfil
&\hfil${#}$
&${#}$
\cr
\Lambda_j
 & = \mu_j\sqrt{G(m_0+m_j)a_j}
  & \xi_j
 & = \sqrt{2\Lambda_j}\sqrt{1-\sqrt{1-e_j^2}} \cos\omega_j\>,
\cr
\lambda_j
   & = \ell_j + \omega_j
  & \eta_j
   & = -\sqrt{2\Lambda_j}\sqrt{1-\sqrt{1-e_j^2}} \sin\omega_j\>,
\cr
}}
$$
dove $m_0,\,m_1,\,m_2$ sono le masse dei tre corpi, $\mu_j = \frac{m_0
m_j}{m_0+m_j}$ con $j=1,2$ le masse ridotte dei due pianeti e $G$ \`e
la costante di gravitazione.  Qui abbiamo omesso le variabili che
descrivono le inclinazioni ed i nodi perch\'e vengono eliminate dalla
riduzione del momento angolare.  Ricordiamo che gli angoli $\lambda$ e
le azioni ad essi coniugate $\Lambda$ vengono detti {\corsivo
variabili\ veloci}, in quanto le frequenze corrispondenti sono quelle
del moto kepleriano, mentre le $\xi_j,\,\eta_j$, che si riferiscono
alle eccentricit\`a, sono dette {\corsivo variabili\ lente} o
{\corsivo secolari}, in quanto nell\:approssimazione kepleriana esse
restano costanti mentre nelle approssimazioni successive sono soggette
a variazioni molto lente visibili solo sull\:arco di secoli.

L\:Hamiltoniana in variabili di Poincar\'e comprende due contributi
\begin{equation}
H = H_0(\Lambda) + H_1(\Lambda,\lambda,\xi,\eta)\ ,
\label{frm:1}
\end{equation}
che consistono in una parte imperturbata $H_0(\Lambda)$ che descrive
il moto kepleriano (ellittico o circolare) ed una perturbazione
$H_1(\Lambda,\lambda,\xi,\eta)$ dovuta all\:interazione mutua tra i
pianeti.  L\:Hamiltoniana cos\`{\i} ottenuta pu\`o essere sviluppata
in serie di potenze intorno alle orbite kepleriane circolari.
A tal fine si sceglie il valore $\Lambda^*$ risolvendo l'equazione
$$
\left.\frac{\partial{\langle H \rangle_{\lambda}}}{\partial\Lambda_{j}}
\right|_{\substack{\Lambda=\Lambda^*\\\xi=\eta=0}} =
n^*_j\ ,\qquad j=1,2\ ,
$$
dove il simbolo $\langle\cdot\rangle_{\lambda}$ indica la media
rispetto agli angoli veloci e $n^*_j$ \`e la frequenza fondamentale
del moto medio relativa all'angolo $\lambda_j$, per una esposizione
pi\`u dettagliata si veda~\dbiref{Loc-Gio-00}.  Si procede poi ad uno
sviluppo in serie di potenze nelle variabili $\xi,\,\eta$
nell\:intorno dell\:origine, osservando che $\xi=\eta=0$ corrisponde
ad eccentricit\`a nulla.  Definendo $\Lambda=\Lambda^*+\Lambda'$
sviluppiamo in serie di potenze di $\Lambda'$
l'Hamiltoniana~(\ref{frm:1}) ed eliminando gli apici per semplificare
le notazioni la riscriviamo nella forma
\begin{equation}
H_0(\Lambda) = \sum_{i} \nu_i\Lambda_i + O(\Lambda^2)\ ,
\label{frm:2}
\end{equation}
che descrive il moto circolare con frequenze kepleriane $\nu$, e
\begin{equation}
H_1 = \sum_{j,k\in\interi^n} c_{j,k}(\Lambda,\lambda) 
       \xi^j \eta^k\ ,
\label{frm:3}
\end{equation}
che \`e lo sviluppo della perturbazione in serie di potenze di
$\xi,\eta$ con coefficienti $c_{j,k}(\Lambda,\lambda)$ periodici in
$\lambda$. 

Lo schema generale esposto fin qui non differisce di molto da quello
di Lagrange, ma nel nostro calcolo introduciamo due differenze
significative.  La prima \`e che, grazie all\:uso della manipolazione
algebrica al calcolatore, siamo in grado di calcolare esplicitamente
anche molti termini di secondo ordine nelle masse e termini non
lineari nelle variabili lente. La seconda differenza \`e che possiamo
garantire un\:approssimazione migliore per la dinamica dei semiassi
maggiori, proprio tenendo conto della perturbazione fino all\:ordine 2
nelle masse.  A tal fine riordiniamo lo sviluppo della perturbazione
nella formula~(\ref{frm:3}) come
\begin{equation}
H_1(\Lambda,\lambda,\xi,\eta) = f_0(\lambda,\xi,\eta) 
  + f_1(\Lambda,\lambda,\xi,\eta) + O(\Lambda^2)\ ,
\label{frm:4}
\end{equation}
dove si intende che $f_0(\lambda,\xi,\eta)$ e
$f_1(\Lambda,\lambda,\xi,\eta)$ sono rispettivamente di grado zero e
uno nelle azioni veloci $\Lambda$, e con una coppia di trasformazioni
canoniche cerchiamo di rimuovere $f_0$ ed $f_1$ in modo da lasciare
tra i contributi che dipendono effettivamente da $\lambda$ solo
quelli che sono almeno di ordine 2 nelle masse.  Questo procedimento
si ispira alla costruzione della forma normale di Kolmogorov, ed \`e
descritto in dettaglio in~\dbiref{Loc-Gio-05}.  Ne risulta
un\:Hamiltoniana che ha ancora la forma~(\ref{frm:3}), ma la
perturbazione $H_1$ non contiene pi\`u alcun termine effettivamente
dipendente da $\lambda$ che sia anche indipendente da o lineare in
$\Lambda$ e di ordine inferiore a 2 nelle masse.

Infine, riprendendo lo schema di Lagrange, introduciamo il {\corsivo
modello secolare}.  A tal fine, con un\:operazione di media,
eliminiamo dall\:Hamiltoniana la dipendenza dagli angoli veloci
$\lambda$ e fissiamo $\Lambda$.  Concretamente ci\`o si ottiene eliminando
dagli sviluppi in serie di Fourier tutti i termini che contengono le
variabili $\lambda$ e ponendo $\Lambda=0$.  Ci\`o corrisponde a
fissare la dinamica dei semiassi maggiori in modo che sia una piccola
variazione quasi periodica dell\:orbita circolare.  In tal modo
l\:Hamiltoniana risultante dipende solo dalle variabili lente
$\xi,\eta$, e si riduce ad un sistema a due gradi di libert\`a.  Il
fatto rilevante \`e che lo sviluppo in serie di potenze
dell\:Hamiltoniana non contiene contributi di grado dispari, ed in
particolare neppure termini lineari, sicch\'e si deve studiare la
dinamica di un sistema conservativo in prossimit\`a di un equilibrio.
Precisamente si ottiene un\:Hamiltoniana della forma
\begin{equation}
H(\xi,\eta) = H_0(\xi,\eta) + H_2(\xi,\eta) +\ldots\ ,
\label{frm:5}
\end{equation}
 dove $H_0,\,H_2,\,\ldots$ sono polinomi omogenei di grado
rispettivamente $2,\,4,\,\ldots$.  

La propriet\`a che abbiamo appena enunciato \`e a prima vista
sorprendente, ed in effetti a Lagrange spetta il merito di averla
messa in evidenza per primo e di aver fondato su di essa le sue
ricerche sui moti secolari.  Il procedimento da lui seguito pu\`o
riformularsi dicendo che si considera la sola parte quadratica
dell\:Hamiltoniana, sicch\'e si deve studiare un sistema di equazioni
lineari a coefficienti costanti.  Il metodo per risolvere tali
equazioni era ben noto a Lagrange, dato che egli stesso lo aveva
sviluppato in una memoria del 1763 dandogli sostanzialmente la forma
che ancora troviamo nei trattati di Analisi Matematica.

Il suo risultato di stabilit\`a consiste poi nel mostrare che le
soluzioni scritte come composizione di moti periodici con le frequenze
e le ampiezza calcolate per i pianeti restano sempre limitate in un
intorno abbastanza piccolo dell\:origine.  La conclusione si fonda
sull\:assunzione, non dimostrata ma accettata come perfettamente
plausibile, che la dinamica nell\:intorno dell\:equilibrio non venga
influenzata in modo rilevante dai contributi non lineari che compaiono
nelle equazioni.  Proprio questo invece \`e il punto che dobbiamo
rimettere in discussione alla luce degli sviluppi delle nostre
conoscenze dopo la scoperta dei moti caotici da parte di Poincar\'e.

Torniamo dunque a considerare l\:Hamiltoniana~(\ref{frm:5}). Grazie
all\:analiticit\`a di tutte le funzioni e le trasformazione coinvolte,
fin qui lo sviluppo risulta essere convergente in un intorno
dell\:equilibrio.  Inoltre, avendo determinato le frequenze $\omega_j$
dei moti secolari col metodo di Lagrange, possiamo anche introdurre
una trasformazione lineare di coordinate che pone la parte quadratica
dell\:Hamiltoniana nella forma diagonale
\begin{equation}
H_0(\xi,\eta) = 
 \sum_{j} \frac{\omega_j}{2} \bigl(\xi_j^2 + \eta_j^2\bigr)\ ,
\label{frm:6}
\end{equation}

\section{La stabilit\`a esponenziale}\label{sec:3}
A questo punto ha inizio la parte pi\`u rilevante della nostra
estensione del lavoro di Lagrange, in quanto teniamo conto proprio dei
contributi non lineari all\:Hamiltoniana~(\ref{frm:5}) che abbiamo
potuto calcolare grazie alla manipolazione algebrica con uso del
calcolatore.

Si fa ricorso alla forma normale di Birkhoff nell\:intorno
dell\:equilibrio.   Si considera un\:Hamiltoniana della forma
\begin{equation}
H(x,y) = H_0(x,y) + H_1(x,y) + H_2(x,y) + \ldots\ ,
\label{frm:7}
\end{equation}
dove
$$
H_0(x,y) = \sum_{j} \frac{\omega_j}{2}(x_j^2 + y_j^2)
$$
e $H_1(x,y),\,H_2(x,y)\ldots$ sono polinomi omogenei di grado
$3,\,4,\ldots\,$, sicch\'e si ha una serie di potenze convergente in
un intorno dell\:origine.  L\:Hamiltoniana~(\ref{frm:5}) ha questa
forma, salvo la particolarit\`a di non contenere termini dispari nello
sviluppo.  L\:obiettivo \`e costruire una trasformazione canonica
di coordinate prossima all\:identit\`a che ponga l\:Hamiltoniana nella
forma
\begin{equation}
Z^{(r)}(x,y) = H_0(x,y) + Z_1(x,y) +\ldots+Z_r(x,y) + 
 \Rscr^{(r+1)}(x,y)\ ,
\label{frm:8}
\end{equation}
dove $Z_1(x,y),\ldots,Z_r(x,y)$ dipendono solo dalle azioni $I_j =
(x_j^2+y_j^2)/2$, e $\Rscr^{(r+1)}(x,y)$ \`e un resto non normalizzato
di grado almeno $r+3$ nelle variabili $x,\,y$.  A tal fine utilizziamo
una successione di trasformazioni canoniche generate mediante
l\:algoritmo della serie di Lie.\footnote{Per un\:esposizione del
metodo delle serie di Lie in ambito hamiltoniano si veda ad
esempio~\dbiref{Giorgilli-95}.}  Precisamente, supponendo di aver
costruito la forma normale $Z^{(r-1)}$ fino ad un ordine $r-1$, si
determina una funzione generatrice $\chi_r(x,y)$ come polinomio
omogeneo di grado $r+2$ risolvendo l\:equazione
\begin{equation}
L_{H_0} \chi_r - Z_r = Q_{r}\ ,
\label{frm:9}
\end{equation}
dove $Q_r$ \`e la parte omogenea di grado $r$ del resto non ancora
normalizzato, e $L_{f}\cdot = \{f,\cdot\}$ \`e la parentesi di Poisson
con la funzione $f$, ovvero la derivata di Lie lungo il campo
hamiltoniano generato da $f$.  Si determina poi la nuova Hamiltoniana
calcolando
$$
Z^{(r)} = \exp\bigl(L_{\chi_r}\bigr) Z^{(r-1)}\ ,
$$
dove 
$$
\exp\bigl(L_{\chi_r}\bigr) = 1 + L_{\chi_r} + 
  \frac{1}{2!} L_{\chi_r}^2 +   \frac{1}{3!} L_{\chi_r}^3 + \ldots\ ,
$$
\`e l\:operatore esponenziale della serie di Lie.  Lo schema di
calcolo \`e facilmente programmabile mediante un manipolatore
algebrico, in quanto richiede solo il calcolo di somme, prodotti e
derivate di polinomi omogenei.

Se, ignorando per un momento il problema della convergenza del
procedimento di costruzione della forma normale, si immagina di
applicare infinite volte lo schema appena descritto si ottiene una
forma normale
$$
Z^{(\infty)}(x,y) = H_0(I) + Z_1(I) +Z_2(I) + \ldots\ ,
$$
funzione solo delle azioni $I_j=(x_j^2+y_j^2)/2$
che risultano essere costanti del moto per $Z^{(\infty)}$.
Scrivendo le equazioni di Hamilton si ottiene cos\`{\i}
$$
\dot x_j = \Omega_j(I) y_j\ ,\quad \dot y_j = -\Omega_j(I) x_j\ ,
$$
dove
$$
\Omega_j(I) = \omega_j + \frac{\partial Z_1}{\partial I_j}(I)
 + \frac{\partial Z_2}{\partial I_j}(I) +\ldots\ ,
$$
sono le frequenze che sono costanti del moto, essendo funzioni solo
delle $I$, e in conseguenza della non linearit\`a delle equazioni
dipendono dal valore iniziale delle azioni $I$.  Le equazioni sono
integrabili in modo elementare, in quanto le soluzioni sono
oscillazioni con frequenze $\Omega(I)$ dipendenti dal dato iniziale.
Se cos\`{\i} fosse potremmo affermare di aver esteso la teoria di
Lagrange nel senso che abbiamo calcolato dei valori migliori per le
frequenze secolari, mantenendo poi tutte le sue conclusioni per quanto
riguarda la stabilit\`a.  Inoltre si giustificherebbe la validit\`a
della teoria di Lagrange in quanto per piccole ampiezze le correzioni
non lineari alle frequenze sono piccole.

Vi sono per\`o due difficolt\`a.  La prima, nota come problema dei
{\corsivo piccoli divisori}, \`e che la soluzione
dell\:equazione~(\ref{frm:9}) \`e possibile solo assumendo delle
condizioni di non risonanza sulle frequenze $\omega$ del moto
imperturbato. Precisamente, dal punto di vista teorico si chiede che
la quantit\`a $\sum_{j} k_j\omega_j$ con coefficienti $k_j$ interi si
annulli solo se $k_j=0$ per tutti i $j$. La seconda difficolt\`a si
cela nella falsit\`a dell\:ipotesi che la forma normale risulti essere
convergente.  Non \`e difficile verificare che ciascun passo del
procedimento \`e ben definito, nel senso che la funzione $Z^{(r)}$,
per ogni $r$ finito, risulta essere convergente in un intorno
dell\:origine, ad esempio una sfera di raggio $\rho_r$.  Le stime
analitiche per\`o consentono solo di dimostrare che $\rho_r$ \`e
limitato inferiormente da una successione che tende a zero almeno come
$1/r$.  Questo non esclude che si possa avere convergenza in casi
specifici, ed in effetti si possono costruire esempi di Hamiltoniane
che ammettono una forma normale di Birkhoff convergente.  Tuttavia nel
lavoro di Siegel~\dbiref{Siegel-41} si mostra che la divergenza \`e il
caso tipico.\footnote{Uno studio numerico che illustra i meccanismi
che conducono alla divergenza si trova
in~\dbiref{Giorgilli-2003},\dbiref{Giorgilli-2004}.}

Ci\`o che rende utile lo sviluppo perturbativo nonostante la
divergenza \`e il carattere asintotico delle serie.  Consideriamo un
intorno $\Delta_{\rho}$ dell\:origine  imponendo la
condizione $\bigl|I_j(x,y)\bigr|\le\rho^2/2$ per $j=1,\ldots,n\,$.
Assumiamo poi che le frequenze $\omega$ soddisfino
la {\corsivo condizione diofantea} $\Bigl|\sum_{j}
k_j\omega_j\Bigr|\ge\gamma|k|^{-\tau}$ con $\gamma\gt 0$ e $\tau\gt
n-1$, essendo $n$ il numero di gradi di libert\`a, e
$|k|=\sum_{j}|k_j|$.  Allora con considerazioni teoriche si pu\`o
verificare che nell\:intorno considerato si ha
$$
\sup_{(x,y)\in\Delta_{\rho}}\bigl|\Rscr^{(r+1)}\bigr| 
 \le B^r (r!)^{n+1} \rho^{r+1}\ ,
$$ 
dove $B$ \`e una costante positiva e $n$ \`e il numero di gradi di
libert\`a del sistema (si veda ad esempio~\dbiref{Giorgilli-88.4}
o~\dbiref{Giorgilli-89}).  Ci\`o mette in evidenza il carattere
asintotico delle serie che stiamo considerando.  Qui l\:ordine $r$ di
normalizzazione \`e arbitrario, ma lo si pu\`o determinare come
funzione $r = r_{\rm opt}(\rho)$  minimizzando la
funzione $(r!)^{n+1}\rho^{r}$.  Si ha cos\`{\i} $r\simeq
(1/\rho)^{1/(n+1)}$, ed utilizzando la formula di Stirling si
valuta
$$
\bigl|\dot I\bigr| \simeq  \exp\bigl(-(1/\rho)^{1/(n+1)}\bigr)\ .
$$
Assumendo che il dato iniziale sia contenuto in un intorno
$\Delta_{\rho/2}$ dell\:origine si pu\`o allora garantire che
l\:orbita rester\`a confinata nel polidisco $\Delta_{\rho}$ per un
tempo $T\simeq\exp\bigl((1/\rho)^{1/(n+1)}\bigr)$ che cresce pi\`u
rapidamente di qualunque potenza al decrescere di $\rho$.  \`E questo,
in forma sintetica, l\:argomento che conduce alla stima esponenziale
del tempo di stabilit\`a tipica della teoria alla
Nekhoroshev~\dbiref{Nekhoroshev-77},\dbiref{Nekhoroshev-79}.

\section{Calcolo effettivo del tempo di stabilit\`a}\label{sec:4}
Ci poniamo ora l\:obiettivo di tradurre l\:argomento che abbiamo
appena esposto in un algoritmo di calcolo che ci consenta di dare una
valutazione esplicita del tempo di stabilit\`a per un sistema reale.
Consideriamo un sistema ad $n$ gradi di libert\`a descritto da
un\:Hamiltoniana della forma~(\ref{frm:7}) troncata ad un ordine
$r_{\max}$ che pu\`o scegliersi, ad esempio, compatibile con la
potenza di calcolo disponibile.

Grazie alla manipolazione algebrica costruiamo esplicitamente la forma
normale di Birkhoff per il nostro sistema fino all\:ordine $r_{\rm
max}$.  In questa fase del calcolo potrebbe presentarsi il problema
dei piccoli divisori, ma possiamo osservare che grazie al troncamento
la condizione di non risonanza deve essere verificata solo per $|k|\le
r_{\rm max}$; questa condizione \`e facile da verificare dal momento che si
deve considerare un numero finito di casi.  Nel calcolare la
forma normale avremo cura anche di tener memoria del primo termine del
resto, ad ogni ordine.  In altre parole, per $r=1,\ldots,r_{\rm max}$
pari costruiamo esplicitamente un\:Hamiltoniana
$$
Z^{(r)} = H_0(I) + Z_1(I) + \ldots + Z^{(r)}(I) 
 + F^{(r+1)}(x,y) + \ldots + F^{(r_{\rm max})}(x,y)\ ,
$$
dove le funzioni $F$ denotano la parte non ancora normalizzata.
Di fatto la condizione di non risonanza implica che le funzioni $Z_j$
si annullino per $j$ dispari, ma ci\`o non ha grande rilevanza
per la discussione di questo paragrafo.

Consideriamo poi un intorno dell\:origine a forma di polidisco con
raggi $R_1,\ldots,R_n$, ossia
$$
\Delta_{\rho R} = \bigl\{(x,y)\in\reali^{2n}\>:\>
  x_j^2+y_j^2 \le \rho^2R_j^2\,,\>1\le j\le n\bigr\}\ .
$$
Scriviamo un polinomio generico di grado $s$ come
$$
f(x,y) = \sum_{j,k} f_{j,k} x^j y^k\ ,\quad |j|+|k| = s\ ,
$$ 
dove abbiamo usato la notazione multiindice $j=(j_1,\ldots,j_n)$,
$k=(k_1,\ldots,k_n)$ e $x^jy^k = x_1^{j_1}\cdot\ldots\cdot y_n^{k_k}$.
Scegliamo $n$ parametri positivi $R=(R_1,\ldots,R_n)$ e calcoliamo la
quantit\`a
\begin{equation}
|f|_R = \sum_{j,k} |f_{j,k}| R^{j+k} \Theta_{j,k}\ ,\quad
\Theta_{j,k} = \sqrt{\frac{j^j k^k}{(j+k)^{j+k}}}\ .
\label{frm:20}
\end{equation}
In tal modo per  $\rho\gt 0$ assegnato abbiamo 
$$
\sup_{(x,y)\in\Delta_{\rho R}} \bigl|f(x,y)\bigr| 
 \lt \rho^s |f|_R\ .
$$
Questa stima richiede qualche giustificazione.  Se consideriamo un
disco di raggio $R_i$ nel piano $x_i,\,y_i$ verifichiamo subito che si
ha $|x_i^jy_i^k| \le R_i^{j+k}\Theta_{j,k}$. Basta infatti scrivere
$x_i=R_i\cos\theta\,,\>y_i=R_i\sin\theta$ e verificare che
sull\:intervallo $0\le\theta\le 2\pi$ si ha
$|\cos^j\theta\sin^k\theta|\le\Theta_{j,k}$.  La quantit\`a $|f|_R$
definita dalla~(\ref{frm:20}) \`e la somma di tutti questi contributi.

Dobbiamo ora valutare $\sup_{(x,y)\in\Delta_{\rho R}}
\bigl|\dot I(x,y)\bigr|$.   Ricordando che la derivata
temporale di una funzione \`e la parentesi di Poisson con
l\:Hamiltoniana facciamo uso della diseguaglianza
\begin{equation}
\bigl|\dot I_j(x,y)\bigr| 
 \lt C\rho^{r+3}\bigl| \{I_j, F^{(r+1)}\}\bigr|_R\ ,
\label{frm:21}
\end{equation}
avendo scelto una costante $C\gt 1$ opportuna.  Qui \`e necessaria
qualche precisazione perch\'e in linea di principio dovremmo tener
conto di una serie infinita, il che \`e chiaramente impossibile in
pratica.  L\:argomento \`e il seguente.  Dalle stime teoriche sappiamo
che la serie dei resti \`e stimata da una serie geometrica.  Se $\rho$
\`e inferiore al raggio di convergenza della forma normale all\:ordine $r$
allora esiste una costante $C$ per cui vale la stima riportata sopra.
Nel calcolo pratico, dal momento che consideriamo $\rho$ abbastanza
piccolo, sceglieremo $C=2$.  Possiamo per\`o osservare che in pratica
la dipendenza del risultato finale dalla scelta di $C$ risulta essere
molto debole.

Veniamo dunque al tempo di stabilit\`a.  Osservando che $I_j \le
\rho^2R_j^2/2$ abbiamo anche $\dot I_j = R_j^2\rho\dot\rho$, e possiamo
riscrivere la diseguaglianza~(\ref{frm:21}) come
$$
\dot\rho \le \frac{B_{r,j}}{R_j^2} \rho^{r+2}\ ,\quad
 B_{r,j} = C \bigl| \{I_j, F^{(r+1)}\}\bigr|_R\ .
$$
Da qui possiamo ricavare una maggiorazione per la funzione $\rho(t)$
risolvendo l\:equazione differenziale $\dot\rho = B_{r,j}
\rho^{r+2}/ R_j^2$.  Separando le variabili otteniamo che il tempo necessario
per passare dal valore iniziale $\rho_0$ ad un $\rho\gt 0$ arbitrario
soddisfa la diseguaglianza $|t| \ge \tau(\rho_0,\rho,r)$, dove
$$
\tau(\rho_0,\rho,r) = \min_j
 \frac{R_j^2}{B_{r,j}} \int_{\rho_0}^{\rho} \frac{d\sigma}{\sigma^r} = 
  \min_j \frac{R_j^2}{(r+1)B_{r,j}}
   \left(\frac{1}{\rho_0^{r+1}} -\frac{1}{\rho^{r+1}}\right)\ .
$$
Qui introduciamo una scelta per $\rho$ ponendo $\rho=2\rho_0$, per cui
la formula precedente diventa
$$
\tau(\rho_0,2\rho_0,r) = \min_j
 \left(1-\frac{1}{2^{r+1}}\right)\frac{R_j^2}{(r+1)B_{r,j}\,\rho_0^{r+1}}\ .
$$
Poich\'e questa stima \`e valida per qualunque scelta di $r$ possiamo
ottimizzarla scegliendo il valore $r_{\rm opt}(\rho_0)$ che massimizza
$\tau(\rho_0,2\rho_0,r)$ ed in tal modo determiniamo un tempo
\begin{equation}
T(\rho_0) = \max_{r}\, \tau(\rho_0,2\rho_0,r)\ .
\label{frm:12}
\end{equation}
Questa \`e la miglior indicazione fornita dal nostro algoritmo, e
chiameremo $T(\rho_0)$ il tempo di stabilit\`a.  Osserviamo che tutte
le quantit\`a scritte sono calcolabili esplicitamente.

\section{Applicazione al sistema Sole--Giove--Saturno}\label{sec:5}
Veniamo infine all\:applicazione al sistema Sole--Giove--Saturno.  La
prima parte del calcolo, del tutto classica, consiste nel calcolare
gli sviluppi in serie di potenze e trigonometriche necessari per dare
una forma esplicita all\:Hamiltoniana del problema dei tre corpi in
variabili di Poincar\'e, nella forma~(\ref{frm:2}) e~(\ref{frm:3}).
Qui scegliamo i valori delle masse e dei parametri orbitali di Giove e
Saturno, riportati per completezza nella
tabella~\ref{tab:parameters_SJS}.   Nello sviluppo teniamo conto
dei contributi fino all\:ordine 2 nelle masse e fino al grado 6 nelle
variabili lente $\xi,\,\eta$. 

Tutto il calcolo \`e stato svolto grazie ad un pacchetto di
manipolazione algebrica realizzato {\corsivo ad hoc} dagli autori, ed
in grado di eseguire tutte le operazioni algebriche necessarie per
l\:applicazione degli algoritmi perturbativi.

\begin{table*}
\caption[]{Masse ed elementi orbitali eliocentrici di Giove e
Saturno calcolati dal JPL per il Giorno Giuliano (JD) $2451220.5\,$.
Le lunghezze sono in Unit\`a Astronomiche (UA); i tempi in anni; la
costante gravitazionale \`e $G=1\,$.  In queste unit\`a la massa del
Sole \`e $4\pi^2\,$.\\ }
\label{tab:parameters_SJS}
\begin{tabular}{|rc|l|l|}
\hline
\phantom{\vbox to 12pt{\relax}}& & Giove ($j=1$) & Saturno ($j=2$)
\\
\hline
\phantom{\vbox to 14pt{\relax}}massa & $m_j$
  & $(4\pi^2)/1047.355$ & $(4\pi^2)/3498.5$
\\
\phantom{\vbox to 12pt{\relax}}semiasse maggiore 
  & $a_j$ & $5.20092253448245$ & $9.55716977296997$
\\
\phantom{\vbox to 12pt{\relax}}anomalia media 
  & $\ell_j$ & $6.14053316064644$ & $5.37386251998842$
\\
\phantom{\vbox to 12pt{\relax}}eccentricit\`a 
  & $e_j$ & $0.04814707261917873$ & $0.05381979488308911$
\\
\phantom{\vbox to 12pt{\relax}}argomento del perielio 
  & $\omega_j$ & $1.18977636117073$ & $5.65165124779163$
\\
\phantom{\vbox to 12pt{\relax}}inclinazione 
  & $i_j$ & $0.006301433258242599$ & $0.01552738031933247$
\\
\phantom{\vbox to 12pt{\relax}}longitudine del nodo 
  & $\Omega_j$ & $3.51164756250381$ & $0.370054908914043$
\\
\hline
\end{tabular}
\end{table*}

Procediamo poi al calcolo della parte secolare del sistema, con
un'approssimazione valida fino all\:ordine 2 nelle masse, seguendo lo schema
illustrato nel paragrafo~\ref{sec:2}.  Otteniamo cos\`{\i} lo sviluppo
dell\:Hamiltoniana secolare nella forma~(\ref{frm:5}) e procediamo al
calcolo delle frequenze ed alla diagonalizzazione della parte
quadratica.  Infine calcoliamo la forma normale di Birkhoff fino
all\:ordine 18, che si \`e rivelato sufficiente per i nostri scopi.

\begin{figure}
\centering
\subfigure[Tempo di stabilit\`a]
{\includegraphics[width=78mm, angle=270]{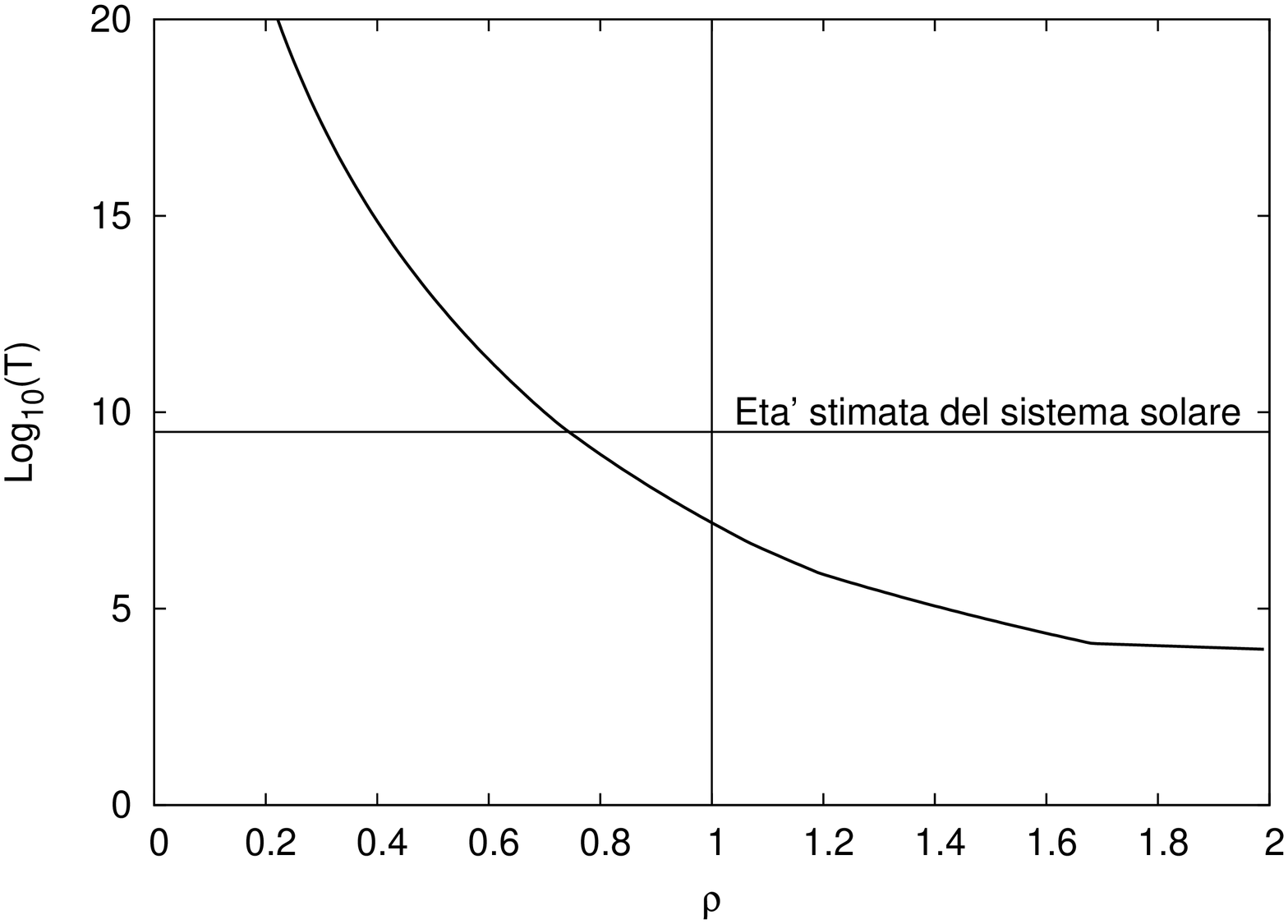}\label{fig:tempi}}
\subfigure[Ordine della normalizzazione ottimale]
{\includegraphics[width=78mm, angle=270]{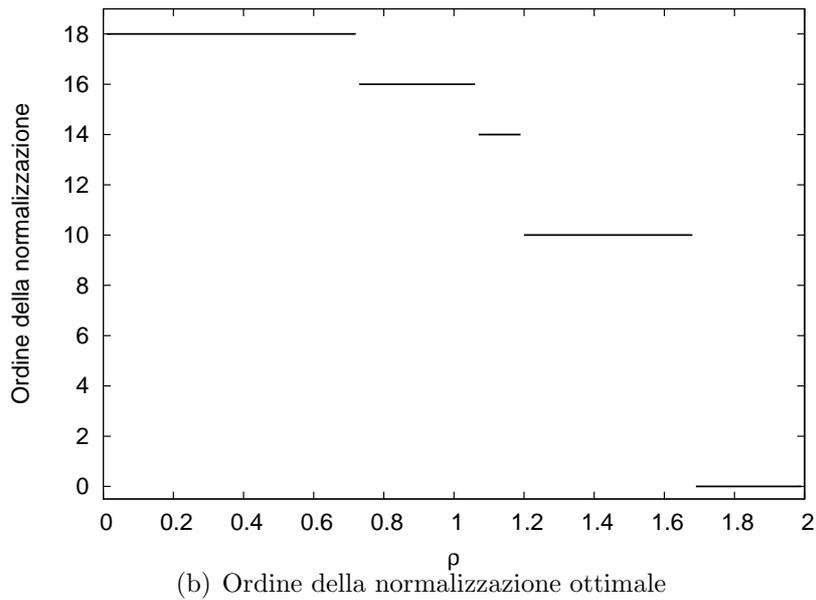}\label{fig:ord_norm}}
\caption{(a)~Stima del tempo di stabilit\`a per il sistema
 Sole--Giove--Saturno al variare del raggio $\rho_0$ del dominio
 contenente i dati iniziali.  I raggi $R$ per il calcolo della norma
 delle funzioni sono scelti in modo che i dati reali per i due pianeti
 corrispondano a $\rho_0=1$, e l\:unit\`a di tempo \`e l\:anno
 terrestre. La scala verticale riporta il logaritmo decimale del
 tempo.  La linea tratteggiata orizzontale indica l\:et\`a stimata del
 sistema solare.  (b)~L\:ordine ottimale di normalizzazione $r_{\rm
 max}$ in funzione di $\rho_0$.}
\end{figure}

Avendo calcolato la forma normale di Birkhoff applichiamo l\:algoritmo
di stima del tempo di stabilit\`a descritto nel paragrafo~\ref{sec:4}.
I risultati sono illustrati nelle figure~\ref{fig:tempi}
e~\ref{fig:ord_norm}.  Nel primo grafico \`e riportato il tempo di
stabilit\`a $T(\rho_0)$ in funzione del raggio $\rho_0$ nel quale sono
contenuti i dati iniziali, in scala semilogaritmica.  Si nota subito
la crescita molto rapida del tempo stimato quando $\rho_0$ decresce.
Che tale crescita sia pi\`u rapida di qualunque potenza lo si arguisce
osservando il secondo grafico, in figura~\ref{fig:ord_norm}, in cui si
riporta l\:ordine ottimale $r_{\rm opt}$ in funzione di $\rho_0$.  I
vari tratti orizzontali corrispondono ad intervalli in cui $r_{\rm
opt}(\rho_0)$ resta costante, ed in ciascuno di questi tratti il tempo
cresce come una potenza $\rho_0^{-r_{\rm opt}}$.  Ma al decrescere di
$\rho_0$ l\:ordine ottimale $r_{\rm opt}$ cresce, ed anche molto
rapidamente: nel nostro grafico il limite $r_{max}=18$ viene raggiunto
per $\rho_0\simeq 0.7$, e poi resta costante solo perch\'e non abbiamo
spinto il calcolo ad ordini pi\`u elevati.

Nel grafico l\:unit\`a di tempo \`e l\:anno terrestre, ed i raggi $R$ sono
scelti in modo che i dati correnti per i due pianeti si trovino nel
polidisco di raggio $\rho_0=1$.  La retta tratteggiata orizzontale
indica l\:et\`a stimata del sistema solare, corrispondente a circa
$5\times 10^9$ anni.  Dalla figura si vede che il tempo di stabilit\`a
stimato col nostro metodo risulta essere di circa $1.5\times 10^7$
anni.

\section{Commenti e possibili sviluppi}\label{sec:6}
La stima che abbiamo ottenuto si rivela ancora pessimistica, in
contrasto ad esempio con le simulazioni numeriche che danno tempi
molto pi\`u lunghi anche per il sistema dei quattro pianeti
maggiori. Possiamo per\`o osservare che non siamo terribilmente
lontani dall\:obiettivo, che sarebbe ragionevole, di raggiungere
almeno l\:et\`a del sistema solare.  Dal grafico, ad esempio, si vede
che basterebbe che l\:eccentricit\`a fosse pari a poco pi\`u di $0.7$ volte
quella reale.  La domanda che si pone spontaneamente \`e se si possano
migliorare i nostri risultati.

Una prima osservazione \`e che lo schema di calcolo che abbiamo
seguito non \`e esente da approssimazioni che possono avere un peso
rilevante.  In effetti l\:algoritmo di stima dei tempi di stabilit\`a
suppone, in buona sostanza, che la perturbazione agisca sempre in modo
da incrementare l\:eccentricit\`a.  Ci\`o \`e certamente falso, ma \`e
difficile tenerne conto nelle stime semianalitiche, mentre le
simulazioni numeriche svolte mediante integrazione diretta delle
equazioni del moto ne tengono conto, di fatto.  In questa luce, il
risultato da noi ottenuto pu\`o gi\`a considerarsi apprezzabile, ed in
effetti si colloca tra i migliori che vengono tipicamente ottenuti
quando si fa ricorso a metodi perturbativi.

Si pone per\`o un problema pi\`u profondo, che chiama in causa in modo
diretto il metodo di Lagrange.  Come abbiamo avuto modo di osservare,
il calcolo delle frequenze secolari \`e stato svolto da Lagrange
facendo riferimento alle orbite circolari.  Se per\`o poniamo il
problema della stabilit\`a su tempi molto lunghi l\:approssimazione
dell\:orbita circolare pu\`o rivelarsi troppo rozza.  In effetti
sappiamo bene che le orbite circolari non sono soluzioni delle
equazioni del problema dei tre corpi, e proprio per questo abbiamo
cercato un\:approssimazione migliore che tenesse conto anche delle
perturbazioni fino al secondo ordine nelle masse.  Il nostro calcolo
mostra che ci\`o non \`e sufficiente: le eccentricit\`a del sistema
reale sono ancora troppo alte.

Ci si chiede allora se si possa migliorare il risultato prendendo come
riferimento delle orbite che abbiano gi\`a eccentricit\`a lontana
dallo zero.  Un tal procedimento in linea di principio \`e possibile.
In effetti abbiamo gi\`a mostrato in una memoria
precedente~\dbiref{Gio-Loc-09} che nelle vicinanze dei dati iniziali
di Giove e Saturno esistono soluzioni quasiperiodiche del tipo
descritto dal teorema di Kolmogorov~\dbiref{Kolmogorov-54}.  Tali
soluzioni possono ben esistere anche se si includono nel modello i
quattro pianeti maggiori, ma dobbiamo ricordare che per sistemi a
pi\`u di 2 gradi di libert\`a l\:esistenza di soluzioni
quasiperiodiche non \`e sufficiente a garantire la stabilit\`a. Siamo
quindi portati ad indagare l'esistenza di un intorno dell\:orbita
quasi periodica che sia stabile per tempi molto pi\`u lunghi di quelli
che abbiamo stimato in questa nota.  In effetti tale calcolo \`e stato
svolto nel lavoro~\dbiref{Giorgilli-2009}, ma si scontra con la
difficolt\`a pratica di dover calcolare un numero troppo elevato di
termini e non \`e raggiungibile con la potenza dei calcolatori
attualmente a nostra disposizione.

Non resta quindi che cercare approssimazioni migliori dell\:orbita
circolare, ma non eccessivamente impegnative dal punto di vista del
calcolo.  Questo \`e un problema aperto al quale stiamo dedicando i
nostri studi.

{}

\end{document}